\documentclass{piner}
\usepackage{rotating}
\begin{document}

\shorttitle{Parsec-Scale Jets of TeV Blazars}

\title{The Parsec-Scale Jets of the TeV Blazars H\,1426+428, 1ES\,1959+650, and PKS\,2155$-$304: 2001--2004}

\author{B.~Glenn~Piner\altaffilmark{1}, Niraj~Pant\altaffilmark{1}, and Philip~G.~Edwards\altaffilmark{2}}

\altaffiltext{1}{Department of Physics and Astronomy, Whittier College,
13406 E. Philadelphia Street, Whittier, CA 90608; gpiner@whittier.edu}

\altaffiltext{2}{CSIRO Australia Telescope National Facility, Locked Bag 194, Narrabri NSW 2390, Australia; Philip.Edwards@csiro.au} 

\begin{abstract}
  We present Very Long Baseline Array (VLBA) observations of the TeV
  blazars H\,1426+428, 1ES\,1959+650, and PKS\,2155$-$304 obtained during
  the years 2001 through 2004.  We observed H\,1426+428 at four epochs
  at 8\,GHz, and found that its parsec-scale structure consisted of a
  $\sim 17$\,mJy core and a single $\sim 3$\,mJy jet component with an
  apparent speed of $2.09c\pm0.53c$.  The blazar 1ES\,1959+650 was
  observed at three epochs at frequencies of 15 and 22\,GHz.  Spectral
  index information from these dual-frequency observations was used to
  definitively identify the core of the parsec-scale structure.
  PKS\,2155$-$304 was observed at a single epoch at 15\,GHz with
  dual-circular polarization, and we present the first VLBI
  polarimetry image of this source. 
  For 1ES\,1959+650 and PKS\,2155$-$304, the
  current observations are combined with the VLBA observations from
  our earlier paper to yield improved apparent speed measurements for
  these sources with greatly reduced measurement errors.  The new
  apparent speed measured for component C2 in 1ES\,1959+650 is
  $0.00c\pm0.04c$ (stationary), and the new apparent speed measured
  for component C1 in PKS\,2155$-$304 is $0.93c\pm0.31c$.  
  We combine the new apparent speed measurements from this paper
  with the apparent speeds measured in TeV blazar jets from our earlier papers to form 
  a current set of apparent speed measurements in TeV HBLs.
  The mean peak apparent pattern speed in the jets of the TeV HBLs is about 1$c$. We conclude
  the paper with a detailed discussion of the interpretation of the
  collected VLBA data on TeV blazars in the context of current
  theoretical models for the parsec-scale structure of TeV blazar
  jets.
\end{abstract}


\section{Introduction}
\label{intro}

The field of TeV gamma-ray astronomy has grown rapidly over the past
several years with the advent of sensitive new gamma-ray
telescopes such as H.E.S.S.\ and MAGIC.  Particularly interesting
results have been obtained in blazar astronomy, as the total number of
detected TeV blazars has increased from six only a few years ago to 17
this year (Wagner 2007b).  The majority of the detected TeV blazars
(16 of 17) belong to the class of high-frequency peaked BL Lac
objects, or HBLs --- so named because the two peaks in their spectral
energy distribution (SED) occur at relatively high UV/X-ray and
GeV/TeV energies. This two-peaked spectral energy distribution in HBLs
is most commonly interpreted as the result of relativistic electrons
(and possibly positrons) radiating in a jet which is undergoing bulk
relativistic motion at a small angle to the observer's line of sight,
with the low-frequency peak due to synchrotron radiation, and the
high-frequency peak due to inverse-Compton scattering of the jet's own
synchrotron photons (synchrotron self-Compton, or SSC emission).  The
TeV source list is mostly limited to relatively nearby blazars ($z\lesssim0.2$),
because of the absorption of TeV gamma-rays on the extragalactic
background light.

The TeV observations have shown dramatic variability in a number of
these blazars. Among the most remarkable variability events are the
200\,s variability timescale detected for PKS 2155$-$304 in July 2006
by H.E.S.S.\ (Aharonian et al.\ 2007), and the 3-minute variations
detected for Mkn\,501 in June 2005 by MAGIC (Wagner 2007a).  Such
rapid variations suggest extremely small emitting volumes and/or time
compression by large relativistic Doppler factors, e.g., $\delta\gtrsim100$
for PKS 2155$-$304 (Aharonian et al.\ 2007).  
High Doppler factors are also sometimes invoked in specific SSC models
of TeV blazar spectra and variability; for example, Fossati et al. (2007)
consider two SSC models to explain the X-ray/TeV variability of Mkn 421:
one with $\delta\sim20$ (scattering in the Klein-Nishina regime), and
one with $\delta\sim100$ (scattering in the Thomson regime).
Such high Doppler factors are
at the upper limit of what is expected in relativistic jets, and
challenge our understanding of these objects if they do in fact occur.

Complementary observations that are crucial to unraveling the physics
of TeV blazar jets are provided by the VLBI technique, which yields
radio images of the relativistic jets with sub-parsec resolutions for
these nearby blazars.  Apparent jet speeds, brightness temperatures,
and limits on jet/counterjet brightness ratios can all be measured
from VLBI images, and these quantities all provide constraints on
fundamental jet parameters such as the bulk Lorentz factor and the
angle of the jet to the line-of-sight (subject to some caveats
discussed at length in $\S$~\ref{disc}).  Our understanding of the
parsec-scale radio properties of the HBL class in general has been
increased by the recent work of Giroletti et al.\ (2004a, 2006). Those
authors studied a sample of low-redshift BL Lac objects with a variety
of radio instruments, and demonstrated that the radio properties of
the HBLs are consistent with them being the beamed versions of nearby
low-luminosity FR I radio galaxies.  The TeV blazars thus likely have
an intrinsically different parent population with weaker jets,
compared to the more distant powerful blazars, which likely have FR II
parents (e.g., Urry \& Padovani 1995).  The mean derived parsec-scale
Lorentz factor for the HBL class by Giroletti et al.\ (2004a),
including TeV sources, is only $<\Gamma>\sim3$, much lower than the
Lorentz factors suggested by the TeV gamma-ray emission.

We have previously published a series of papers investigating the
parsec-scale kinematic properties of TeV-detected HBLs through
multi-epoch high-resolution VLBI observations, predominantly 
with the National Radio
Astronomy Observatory's Very Long Baseline Array (VLBA) \footnote{The
  National Radio Astronomy Observatory is a facility of the National
  Science Foundation operated under cooperative agreement by
  Associated Universities, Inc.}, and we extend that study in this
paper. Previous papers in this series have included: early VLBA and
space VLBI
observations of Mkn\,421 (Piner et al.\ 1999),  
VLBI observations of Mkn\,501 (Edwards \& Piner 2002), 
VLBA observations of 1ES\,1959+650,
PKS\,2155$-$304, and 1ES\,2344+514 (Piner \& Edwards 2004), and new VLBA
polarimetry observations of Mkn\,421 (Piner \& Edwards 2005). This
fifth paper in the series adds new VLBA observations of the TeV
blazars \object{1ES 1959+650} and \object{PKS 2155-304} whose
kinematics were relatively poorly determined in Piner \& Edwards
(2004) (hereafter Paper~I), as well as a four-epoch series of VLBA
observations of \object{H 1426+428} --- the sixth TeV blazar to be
discovered.  Altogether we present six new images of 1ES\,1959+650,
four new images of H\,1426+428, and one new polarization image of
PKS\,2155$-$304, for a total of eleven new datasets. These datasets
comprise the observations from our TeV blazar monitoring program over
the years 2001-2004 (excepting the observations of Mkn\,421, which were
published separately in Piner \& Edwards [2005]).  Note that except
for the two brightest sources (Mkn\,421 and Mkn\,501), the TeV blazars
are too faint in the radio ($\lesssim$100\,mJy) to be included in other
VLBA monitoring programs such as MOJAVE, and they typically require
long observations to obtain images of sufficient dynamic range.
Observational backgrounds on the three specific sources studied in
this paper are presented in the results section for the specific
source.

Our earlier work on the parsec-scale structure of TeV blazars has
shown a noticeable lack of superluminal components in their jets,
which contrasts with the rapid superluminal motions observed in the
jets of more powerful blazars, and with the high Lorentz factors
derived from modeling the TeV emission. 
 We have previously
interpreted the general lack of superluminal components in HBLs
as evidence for a lower bulk Lorentz factor in the
parsec-scale radio-emitting region compared to the TeV-emitting
region.  Models that have been invoked to explain this `bulk Lorentz
factor crisis' include: jets that are decelerated along their length
(Georganopoulos \& Kazanas 2003; Wang, Li, \& Xue 2004; Bicknell et
al.\ 2005), jets with transverse velocity structures consisting of a
fast spine and a slower layer (Giroletti et al.\ 2004b; Ghisellini,
Tavecchio, \& Chiaberge 2005; Henri \& Saug\'{e} 2006), or jets with
opening angles large enough that unintentional averaging over multiple
viewing angles (because of limited resolution) causes the apparent
conflict (Gopal-Krishna, Dhurde, \& Wiita 2004; Gopal-Krishna, Wiita,
\& Dhurde 2006).  Finally, Gopal-Krishna et al.\ (2007) consider a
combination of the last two models; i.e., large opening angle jets with
transverse velocity structures.  Some of these models may be
distinguished through the observed statistical distribution of
apparent speeds in TeV blazar jets (e.g., Gopal-Krishna et al.\ 2006),
so we conclude the paper by presenting our current best set of TeV
blazar apparent speed measurements from the complete series of five
papers, which consists of sixteen component speeds in six sources, and
discussing the various models in the context of the current
observations.

In this paper we use cosmological parameters $H_{0}=71$ km s$^{-1}$
Mpc$^{-1}$, $\Omega_{m}=0.27$, and $\Omega_{\Lambda}=0.73$ (Bennett et
al.\ 2003).  When results from other papers are quoted, they have been
converted to this cosmology.

\section{Observations and Data Reduction}
\label{obs}

We observed H\,1426+428 with the VLBA at 8.4\,GHz at four epochs
between 2001 July and 2003 October, under observation codes BE024 and
BE029.  
A VLBI observation of H\,1426+428 had previously been made at 5\,GHz
by Kollgaard, Gabuzda, \& Feigelson (1996), however our
first observation (BE024 on 2001 July 4) was a pilot study
to see if H\,1426+428 had a parsec-scale jet that could be imaged 
at 8\,GHz with
the VLBA, and to determine a more accurate position for
the source.  That observation was carried out in
phase-referencing mode both to determine a milliarcsecond (mas) level 
position, and because the correlated flux density of
H\,1426+428 was uncertain. The 
source J1419+3821, for which a sub-mas position has been
determined by the International Celestial Reference Frame
(Ma et al.\ 1998), was used as the phase-reference source. 
J1419+3821 has an angular separation of 4.6$^\circ$ from
H\,1426+428.

\begin{table*}[!t]
\caption{Observation Log}
\begin{center}
\label{obstab}
{\scriptsize \begin{tabular}{l l l l c c c} \tableline \tableline \\ [-5pt]
& & & & & Time On & \\
& & \multicolumn{1}{c}{VLBA Observing} & \multicolumn{1}{c}{VLBA} & Frequency & Source & \\
\multicolumn{1}{c}{Source} & \multicolumn{1}{c}{Epoch} & \multicolumn{1}{c}{Code} &
\multicolumn{1}{c}{Antennas$^{a}$} & (GHz) & (hr) & Dual Pol$^{b}$ \\ \tableline \\ [-5pt]
H\,1426+428     & 2001 Jul 4  & BE024  & All   & 8.4  & 7   & No  \\ 
               & 2003 Mar 22 & BE029A & No SC & 8.4  & 7   & No  \\ 
               & 2003 Jul 6  & BE029B & All   & 8.4  & 7   & No  \\
               & 2003 Oct 19 & BE029C & All   & 8.4  & 7   & No  \\ 
1ES\,1959+650   & 2003 Jun 9  & BE030A & No KP & 15.4 & 2.5 & No  \\
               & 2003 Jun 9  & BE030A & No KP & 22.2 & 2.5 & No  \\
               & 2003 Oct 31 & BE030B & All   & 15.4 & 2.5 & No  \\
               & 2003 Oct 31 & BE030B & All   & 22.2 & 2.5 & No  \\
               & 2004 Feb 8  & BE030C & All   & 15.4 & 2.5 & No  \\
               & 2004 Feb 8  & BE030C & All   & 22.2 & 2.5 & No  \\ 
PKS\,2155$-$304 & 2003 Sep 20 & BE031  & No BR & 15.4 & 4   & Yes \\ \tableline
\end{tabular}}
\end{center}
{\scriptsize $a$: BR = Brewster, Washington,
KP = Kitt Peak,Arizona, and SC = Saint Croix, U.S. Virgin Islands.}\\
{\scriptsize $b$: Whether or not the experiment recorded dual-circular polarization.}
\end{table*}

\begin{table*}
\caption{Parameters of the Images}
\begin{center}
\label{imtab}
{\scriptsize \begin{tabular}{l l c l c c l c c} \tableline \tableline \\ [-5pt]
& & & \multicolumn{3}{c}{\underline{Natural Weighting}} & \multicolumn{3}{c}{\underline{Uniform Weighting}} \\ [5pt]
& & & & Peak Flux & Lowest & & Peak Flux & Lowest \\ 
& & Frequency  & & Density & Contour$^{b}$ & & Density & Contour$^{b}$ \\ 
\multicolumn{1}{c}{Source} & \multicolumn{1}{c}{Epoch} & (GHz) & \multicolumn{1}{c}{Beam$^{a}$} &
(mJy beam$^{-1}$) & (mJy beam$^{-1}$) & \multicolumn{1}{c}{Beam$^{a}$}
& (mJy beam$^{-1}$) & (mJy beam$^{-1}$) \\ \tableline \\ [-5pt]
H\,1426+428     & 2001 Jul 4  & 8.4  & 1.54,1.10,$-$4.9  & 20  & 0.13 & 1.07,0.77,$-$7.4  & 18  & 0.28 \\
               & 2003 Mar 22 & 8.4  & 2.17,1.12,$-$19.2 & 19  & 0.12 & 1.63,0.75,$-$20.7 & 19  & 0.22 \\
               & 2003 Jul 6  & 8.4  & 1.80,1.01,$-$7.9  & 16  & 0.11 & 1.30,0.69,$-$8.7  & 15  & 0.20 \\
               & 2003 Oct 19 & 8.4  & 1.77,1.01,$-$5.2  & 16  & 0.11 & 1.26,0.69,$-$4.4  & 15  & 0.19 \\
1ES\,1959+650   & 2003 Jun 9  & 15.4 & 0.97,0.48,$-$4.3  & 107 & 0.37 & 0.74,0.36,$-$2.4  & 98  & 0.67 \\
               & 2003 Oct 31 & 15.4 & 1.01,0.53,$-$6.1  & 104 & 0.35 & 0.75,0.38,$-$5.2  & 95  & 0.55 \\
               & 2004 Feb 8  & 15.4 & 1.12,0.54,5.3     & 77  & 0.36 & 0.77,0.36,1.3     & 67  & 0.70 \\
               & 2003 Jun 9  & 22.2 & 0.65,0.32,$-$1.6  & 101 & 1.09 & 0.52,0.26,0.8     & 92  & 1.68 \\
               & 2003 Oct 31 & 22.2 & 0.72,0.41,$-$1.0  & 84  & 0.78 & 0.53,0.26,$-$0.6  & 76  & 1.38 \\
               & 2004 Feb 8  & 22.2 & 0.85,0.43,10.5    & 61  & 0.72 & 0.56,0.26,10.0    & 52  & 1.41 \\
PKS\,2155$-$304 & 2003 Sep 20 & 15.4 & 1.92,0.51,$-$11.0 & 122 & 0.38 & 1.30,0.36,$-$7.8  & 110 & 0.77 \\ \tableline
\end{tabular}}
\end{center}
{\scriptsize $a$: Numbers given for the beam are the FWHMs of the major
and minor axes in mas, and the position angle of the major axis in degrees.
Position angle is measured from north through east.}\\
{\scriptsize $b$: The lowest contour is set to be three times the rms noise
in the image. Successive contours are each a factor of 2 higher.}
\end{table*}

We were able to model the phase-reference source, J1419+3821,
with two circular Gaussian components: a 0.34\,Jy unresolved core
component and a 0.16\,Jy
component with a FWHM size of 1.2\,mas located 0.6\,mas to the south.
Based on the ICRF position for J1419+3821, we 
derived a (J2000) position for H\,1426+428 of
($\alpha$,$\delta$)=(14h28m32.609s, +42$^\circ$40$'$21.05$''$),
which we expect to have an accuracy of better than 20\,mas.
With the use of this position, 
phase-referencing was unnecessary in subsequent epochs, 
as fringes could be directly detected to H\,1426+428, 
with a correlated flux density of $\sim20$\,mJy at 8\,GHz.
Following the successful imaging of parsec-scale
structure in H\,1426+428 in the pilot study, 
three further epochs were observed throughout
2003 (observation code BE029) to study the jet kinematics.  All four
observations recorded approximately 7 hours on-source, for an expected
thermal noise limit of about 0.05\,mJy beam$^{-1}$. Note that
H\,1426+428 was observed at a lower frequency than the other sources in
this paper and in Paper~I, in order to achieve the needed sensitivity
to image this fainter source.

1ES\,1959+650 was observed with the VLBA during three 6-hour sessions
between 2003 June and 2004 February, under observation code BE030.
The observations were split between 15.4 and 22.2\,GHz, with
approximately 2.5 hours spent on source at each frequency at each
epoch (with approximately one hour of calibration scans and slewing).
This source was observed at two frequencies in order to obtain
spectral index information between the 15 and 22\,GHz images, to help
resolve potential ambiguities in the source structure discussed in Paper~I. The
15\,GHz data has been combined with the three epochs at this frequency
from Paper~I, to obtain kinematic information over a six-epoch series
for this source.

PKS\,2155$-$304 was observed at a single 6-hour epoch on 2003 September
20 at 15.4\,GHz, with dual-circular polarization, under observation
code BE031.  The goal of this observation was two-fold: to provide an
additional epoch several years removed from the three epochs in Paper~I 
to better constrain the jet proper motion (which was poorly
constrained in Paper~I), and to produce the first VLBI polarimetry
image of this source.  Previous observations in Paper~I had shown the
correlated flux density was large enough (at $>$100\,mJy) to produce a
useful polarization image at this frequency. Approximately four hours
were spent on-source in this observation, with the additional two
hours spent on polarization calibration sources.

\begin{table*}[!t]
\caption{Circular Gaussian Models}
\begin{center}
\label{mfittab}
{\scriptsize \begin{tabular}{l l c c c c r c c c} \tableline \tableline \\ [-5pt]
& & Frequency & & $S$ & $r$ &
\multicolumn{1}{c}{PA} & $a$ & & $T_{B}$ \\ 
\multicolumn{1}{c}{Source} & \multicolumn{1}{c}{Epoch} & (GHz) & Component & (mJy) & (mas) &
\multicolumn{1}{c}{(deg)} & (mas) & $\chi_{R}^{2}$ & ($10^{10}$ K) \\ 
\multicolumn{1}{c}{(1)} & \multicolumn{1}{c}{(2)} & (3) & (4) & (5) & (6) &
\multicolumn{1}{c}{(7)} & (8) & (9) & (10) \\ \tableline \\ [-5pt]
H\,1426+428     & 2001 Jul 4  & 8.4  & Core & 19  & ...  & ...     & 0.22 & 0.96 & 0.8  \\ 
               &             &      & C1   & 4   & 1.02 & $-$21.4 & 1.38 &      &      \\
               & 2003 Mar 22 & 8.4  & Core & 19  & ...  & ...     & 0.17 & 0.67 & 1.2  \\ 
               &             &      & C1   & 3   & 1.65 & $-$26.8 & 1.46 &      &      \\
               & 2003 Jul 6  & 8.4  & Core & 16  & ...  & ...     & 0.19 & 0.64 & 0.8  \\ 
               &             &      & C1   & 3   & 1.60 & $-$25.5 & 1.74 &      &      \\
               & 2003 Oct 19 & 8.4  & Core & 15  & ...  & ...     & 0.14 & 0.64 & 1.5  \\ 
               &             &      & C1   & 4   & 1.48 & $-$23.9 & 2.03 &      &      \\
1ES\,1959+650   & 2003 Jun 9  & 15.4 & Core & 81  & ...  & ...     & 0.07 & 0.72 & 10.0 \\
               &             &      & C2   & 72  & 0.34 & 124.0   & 0.55 &      &      \\
               & 2003 Oct 31 & 15.4 & Core & 81  & ...  & ...     & 0.09 & 0.72 & 5.9  \\
               &             &      & C2   & 58  & 0.32 & 125.5   & 0.62 &      &      \\
               & 2004 Feb 8  & 15.4 & Core & 63  & ...  & ...     & 0.15 & 0.73 & 1.5  \\
               &             &      & C2   & 46  & 0.41 & 129.8   & 0.51 &      &      \\
               & 2003 Jun 9  & 22.2 & Core & 90  & ...  & ...     & 0.09 & 0.58 & 2.6  \\
               &             &      & C2   & 72  & 0.33 & 124.2   & 0.48 &      &      \\
               & 2003 Oct 31 & 22.2 & Core & 82  & ...  & ...     & 0.13 & 0.59 & 1.2  \\
               &             &      & C2   & 35  & 0.42 & 126.5   & 0.54 &      &      \\
               & 2004 Feb 8  & 22.2 & Core & 59  & ...  & ...     & 0.15 & 0.55 & 0.7  \\
               &             &      & C2   & 30  & 0.46 & 126.7   & 0.47 &      &      \\
PKS\,2155$-$304 & 2003 Sep 20 & 15.4 & Core & 130 & ...  & ...     & 0.23 & 0.68 & 1.4  \\
               &             &      & C1   & 15  & 1.04 & 163.9   & 0.48 &      &      \\ \tableline
\end{tabular}}
\end{center}
NOTES.--- Col. (5): Flux density in millijanskys.
Col. (6) and (7): $r$ and PA are the polar coordinates of the
center of the component relative to the presumed core.
Position angle is measured from north through east.
Col. (8): $a$ is the Full Width at Half Maximum (FWHM) of the circular Gaussian
component.
Col. (9): The reduced chi-squared of the model fit.
Col. (10): The maximum source-frame brightness temperature of the circular Gaussian core component is given by
$T_{B}=1.22\times10^{12}\;\frac{S(1+z)}{a^{2}\nu^{2}}$~K,
where $S$ is the flux density of the Gaussian in Janskys,
$a$ is the FWHM of the Gaussian in mas,
$\nu$ is the observation frequency in\,GHz, and $z$ is the redshift.
\end{table*}

An observation log of all new observations discussed in this paper is
given in Table~\ref{obstab}.  All observations were recorded at a data
rate of 128 Mbps.  The AIPS software package was used for calibration
and fringe-fitting of the correlated visibilities, and the
visibilities were edited and final CLEAN images were produced using
the Difmap software package. For the polarization experiment BE031,
calibration of the polarization response of the feeds was done with
LPCAL in AIPS.  The required electric vector position angle (EVPA)
correction for BE031 was applied using CLCOR in AIPS, and was
determined from the observed EVPA of calibrator source J2136+006
(which had a relatively stable EVPA during 2003), compared with the
EVPA recorded for this source on the VLA/VLBA Polarization Calibration
Page\footnote{http://www.vla.nrao.edu/astro/calib/polar/},
interpolated to our frequency and epoch of observation.  Parameters
for all images are given in Table~\ref{imtab}; individual images are
discussed in the results sections on the individual sources below.
All data sets are shown restored with both uniform weighting
(uvweight=2,0 in Difmap, improving resolution at the expense of
signal-to-noise ratio) and natural weighting (uvweight=0,$-$2 in
Difmap).  None of the naturally weighted images have an rms noise
level that significantly exceeds the expected thermal noise limit.

\begin{sidewaysfigure*}[t]
\includegraphics[scale=0.80,angle=-90.0]{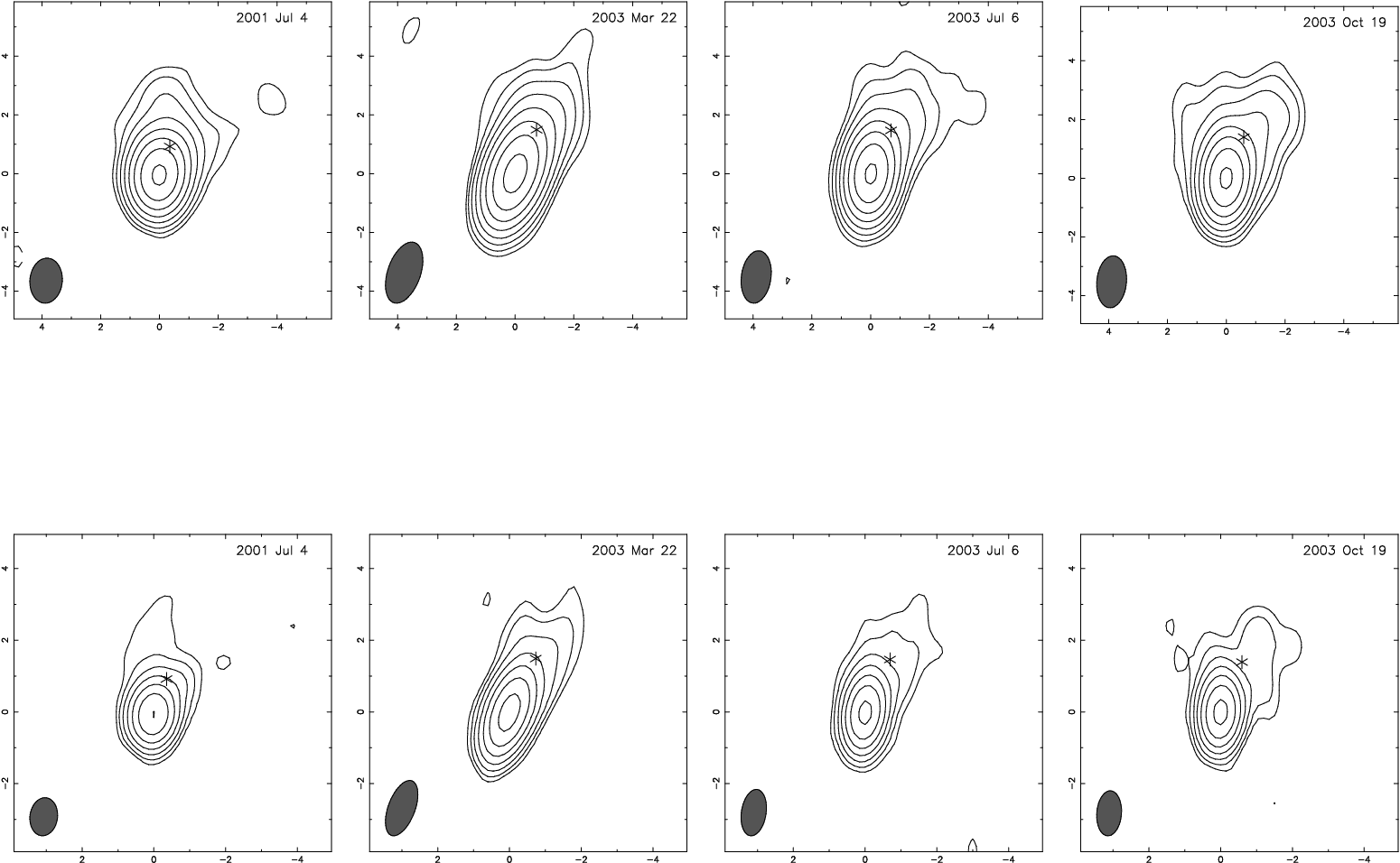}
\caption{VLBA images of H\,1426+428 at 8.4\,GHz.
The top row shows the images obtained with natural weighting,
the bottom row the images obtained with uniform weighting.
The axes are labeled in milliarcseconds (mas).
Numerical parameters of the images are given in Table~\ref{imtab}.
The location of the center of the circular Gaussian (excluding the Gaussian representing the core) 
that was fit to the visibilities is marked with an asterisk.
Parameters of the Gaussian models are given in Table~\ref{mfittab}.}
\end{sidewaysfigure*}

After final calibration of the visibilities, circular Gaussian model
components were fit to the visibilities using the {\em modelfit} task
in Difmap. Such models allow the structure of the jet to be described
using only a few numerical parameters. The source structures in the
images in this paper are relatively simple (partly due to the high
resolution and limited dynamic range of these observations, and some
of these sources show more complex structures on lower-resolution
images that may not be well modeled as circular Gaussians), and in all
cases satisfactory fits to the visibilities were obtained using only
two circular Gaussian components (representing the core and one jet
component).  The model fits are tabulated in Table~\ref{mfittab}, and
results for each source are discussed in the results sections on the
individual sources below.  The reduced $\chi^{2}$ for all model fits
was under 1.0.  Model fitting directly to the visibilities allows
sub-beam resolution to be obtained (e.g., Kovalev et al.\ 2005), and
components can be clearly identified in the model fitting even when
they appear blended with the core component in the CLEAN images.

\begin{figure*}[t]
\begin{center}
\includegraphics[scale=0.35,angle=90.0]{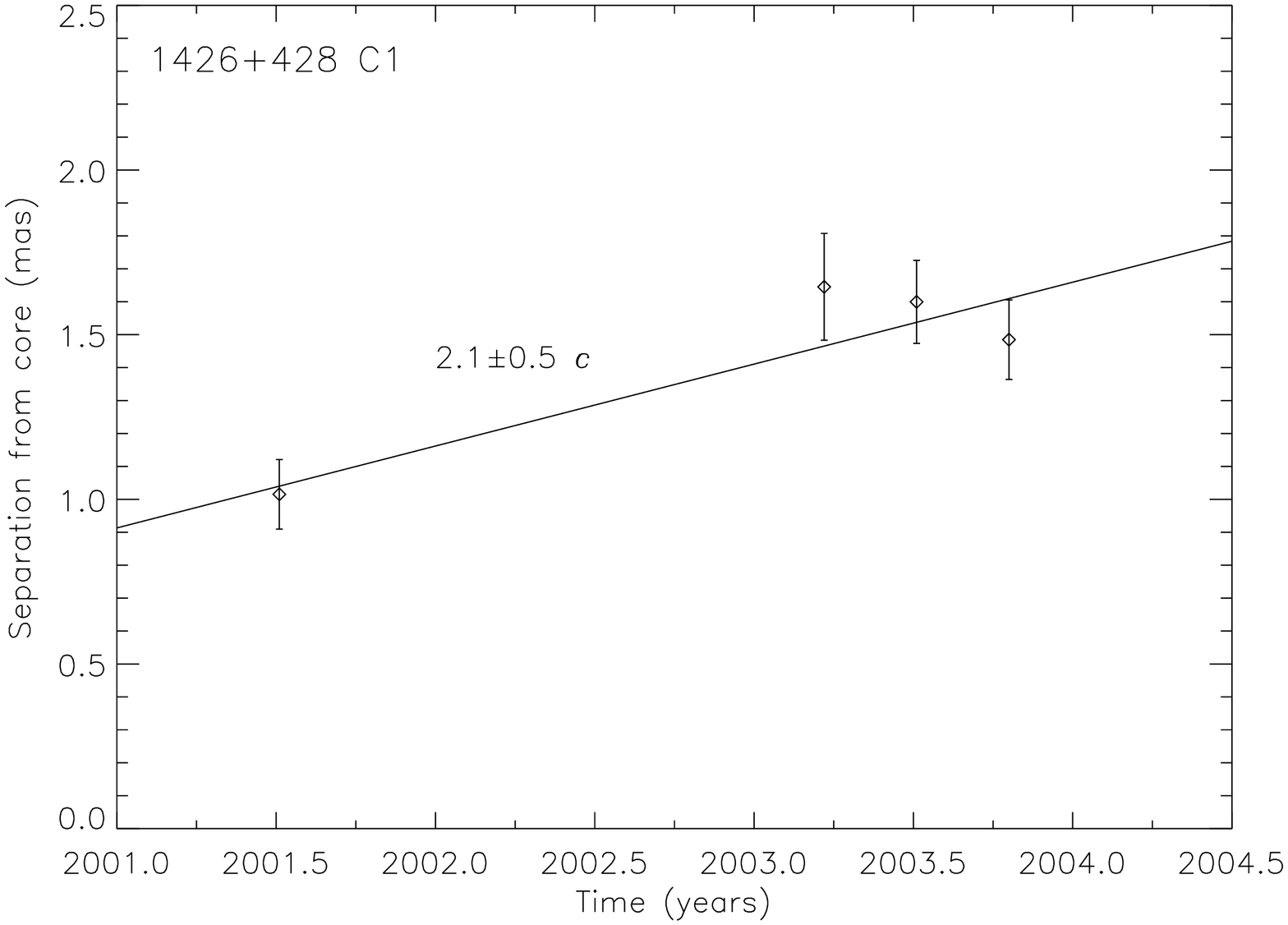}
\caption{Distance from the core of the center of Gaussian component C1 in H\,1426+428 as
a function of time. The line is the least-squares fit to outward motion with constant speed.}
\end{center}
\end{figure*}

As in Paper~I, we estimate that these full-track observations allow us
to fit positions of model component centers to within $\sim10$\% of a
uniform beam width, and this is the error assumed for subsequent
analysis (calculated by taking 10\% of the projection of the
elliptical beam FWHM onto a line joining the center of the core to the
center of the component).  
Error bars much larger than this result in fits to
linear component motion that are so good as to be statistically
unlikely, confirming that this estimated error is reasonable.
Note that the low rms noise achieved by these long
integrations allows us to detect even the faintest component (the 3
mJy component in H\,1426+428) with a signal-to-noise ratio exceeding
30:1.  
Lower limits to the fitted core sizes (which yield upper limits to the
measured brightness temperatures), were determined using the DIFWRAP
program for model component error analysis (Lovell 2000).
These size lower limits range from zero (completely unresolved components)
to only about 20\% less than the best-fit size, and they are discussed in the
text on the core brightness temperature upper limits for each
individual source below.

\section {Results on Individual Sources}
\label{results}

\subsection{H\,1426+428}

The detection in TeV gamma-rays of the HBL H\,1426+428 by the Whipple,
HEGRA, and CAT telescopes (Horan et al.\ 2002; Aharonian et al.\ 2002;
Djannati-Ata{\"i} et al.\ 2002) made it the sixth HBL source to be detected
in high-energy gamma-rays, and, at a redshift of $z=0.129$, it was the
most distant TeV HBL known at that time.  Further analysis of the TeV
gamma-ray emission is given by Petry et al.\ (2002) and Aharonian et
al.\ (2003). Aharonian et al.\ (2003) and Costamante et al.\ (2003)
consider various models for the absorption of gamma-rays by the
extragalactic background light, and find that the data imply an
intrinsic TeV spectrum which dominates the broad-band spectral energy
distribution.  X-ray observations indicate that the synchrotron peak
is sometimes located in excess of 100 keV (Costamante et al.\ 2001;
Wolter et al.\ 2007), and that the location of the peak is
significantly variable (Falcone, Cui, \& Finley 2004).  SSC models
have been applied to the SED of H\,1426+428 by Wolter et al.\ (2007),
Costamante et al.\ (2003) (both using the one-zone homogeneous SSC
model of Ghisellini, Celotti, \& Costamante [2002]), and Kato,
Kusunose, \& Takahara (2006); all of these models give Doppler and
bulk Lorentz factors of about 20, and jet viewing angles of about
2$\arcdeg$, but involve numerous assumptions.

\begin{figure*}[p]
\begin{center}
\includegraphics[scale=0.50,angle=-90.0]{f3.ps}
\caption{VLBA images of 1ES\,1959+650 at 15.4\,GHz.
The top row shows the images obtained with natural weighting,
the bottom row the images obtained with uniform weighting.
The axes are labeled in milliarcseconds (mas).
Numerical parameters of the images are given in Table~\ref{imtab}.
The location of the center of the circular Gaussian (excluding the Gaussian representing the core) 
that was fit to the visibilities is marked with an asterisk.
Parameters of the Gaussian models are given in Table~\ref{mfittab}.}
\end{center}
\end{figure*}

\begin{figure*}
\begin{center}
\includegraphics[scale=0.50,angle=-90.0]{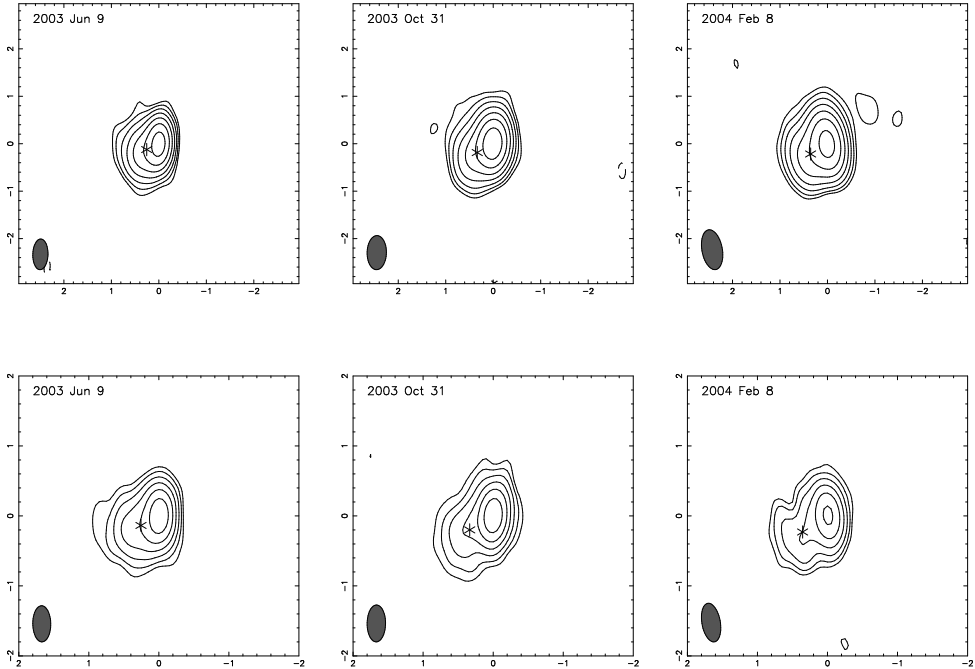}
\caption{VLBA images of 1ES\,1959+650 at 22.2\,GHz.
The top row shows the images obtained with natural weighting,
the bottom row the images obtained with uniform weighting.
The axes are labeled in milliarcseconds (mas).
Numerical parameters of the images are given in Table~\ref{imtab}.
The location of the center of the circular Gaussian (excluding the Gaussian representing the core) 
that was fit to the visibilities is marked with an asterisk.
Parameters of the Gaussian models are given in Table~\ref{mfittab}.}
\end{center}
\end{figure*}

VLA images of this source are shown by Laurent-Muehleisen et al.\ 
(1993) and Giroletti et al.\ (2004a).  Laurent-Muehleisen et al.\ (1993)
present a 1.5\,GHz combined A- and C-configuration image that shows
poorly represented diffuse emission north of the core. Giroletti et
al.\ (2004a) detect the presence of a faint halo surrounding the
central core, with a northeast extension oriented at
P.A.$\sim50\arcdeg$, in their 1.4\,GHz A-array image.  
H\,1426+428 has
been previously observed with VLBI by Kollgaard et al.\
(1996), and by Giroletti et al.\ (2006). 
The 5\,GHz MkIII VLBI image
from 1991 by Kollgaard et al.\ (1996) shows a 2\,mJy component 1.2 mas
northeast (P.A.$\approx20\arcdeg$) of a 19\,mJy compact core.  The
source is completely unresolved in the lower resolution 1.6\,GHz
European VLBI Network (EVN) image in Giroletti et al.\ (2006). Based on
the available VLA and VLBI data, Giroletti et al.\ (2006) estimate a
viewing angle of $\theta\sim20\arcdeg$ and a Doppler factor of
$\delta\sim3$, but those estimates are based on assumptions that may
only be valid in a statistical sense for samples of many blazars (such
as that the Lorentz factor $\Gamma\sim1/\theta$).

\begin{figure*}[t]
\begin{center}
\includegraphics[scale=0.35,angle=0.0]{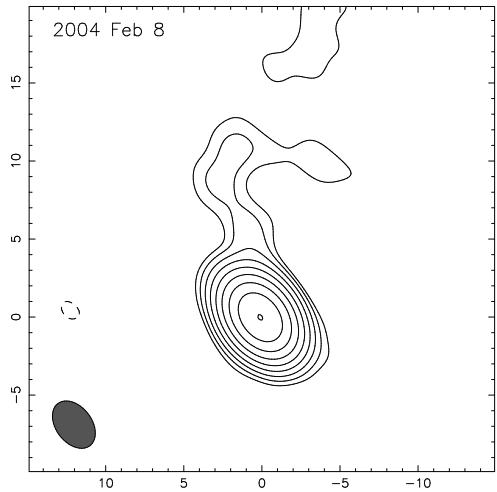}
\end{center}
\caption{Tapered 15\,GHz VLBA image of 1ES\,1959+650 from 2004 Feb 8.
The restoring beam is 3.39 by 2.3 mas at a position angle of 37$\arcdeg$.
The lowest contour is drawn at a flux density of 0.42\,mJy beam$^{-1}$
(three times the rms noise level). Successive contours are each a factor of 2 higher.
The peak flux density is 107\,mJy beam$^{-1}$.}
\end{figure*}

Our four 8.4\,GHz VLBA images of H\,1426+428 are shown in Figure~1.
The top row of Figure~1 shows the naturally weighted images, while the
bottom row shows the uniformly weighted images.  Parameters of the
images are given in Table~\ref{imtab}.  The linear resolution of these
images at the distance of H\,1426+428 is approximately 2.3 pc
mas$^{-1}$.  The images all show a compact core with a flux density of
$\sim$17\,mJy, and a faint jet extending to the northwest, with a
position angle of about $-25\arcdeg$.  The position angle of the
extended structure seen at all four epochs is similar, and differs
significantly from the position angle of the structure measured by
Kollgaard et al.\ (1996) from their single 5\,GHz MkIII VLBI image from
1991 of $\approx20\arcdeg$.  No counterjet is detected; although,
because of the low flux density of the jet, the limit that can be
placed on the jet to counterjet brightness ratio is a rather low 30:1.
Although variability and different resolutions may 
cause significant uncertainty, we can estimate
a total VLBI spectral index between 1.6 GHz (Giroletti et al.\ 2006), 5 GHZ  
(Kollgaard et al.\ 1996) and 8.4 GHz (this paper), and obtain
$\alpha=-0.26+\-0.10$ ($S\propto\nu^{\alpha}$), a flat but not inverted spectral index. 

The structure of H\,1426+428 at all four epochs is well modeled by two
circular Gaussian components, representing the core and a single jet
component, designated C1, with a flux density of about 3\,mJy.
Parameters of the Gaussian models are given in Table~\ref{mfittab}.
The measured brightness temperature of the core is about
$1\times10^{10}$~K (see Table~\ref{mfittab}), but this component is
unresolved (the visibilities are almost as well fit by a point-like
component), so that this brightness temperature should be considered a
lower-limit.  The jet component appears to move out between 2001 and
2003, although no motion can be detected in the three 2003 epochs
alone. A linear fit to the separation of the jet component from the
core vs. time is shown in Figure~2. The fit yields a measured apparent
speed for the jet component of $2.09c\pm0.53c$.  Interpretation of
model fitting data over only a few widely spaced epochs can lead to
uncertainties in component identification (e.g., Piner et al.\ 2007),
but for this source the structure is quite simple, minimizing the
chances of confusing multiple components.  If this measured apparent
speed is taken as the bulk apparent speed of the 
jet\footnote {the apparent jet speed
  $\beta_{app}=\beta\sin\theta/(1-\beta\cos\theta)$, and the Doppler
  factor $\delta=1/\Gamma(1-\beta\cos\theta)$, where $\Gamma$ is the
  bulk Lorentz factor, $\beta=v/c$, and $\theta$ is the angle to the
  line of sight}, 
and is combined with the viewing angle derived from SSC modeling of
about 2$\arcdeg$ (see above), then the bulk Lorentz factor of the jet
at the location of C1 is $\Gamma=5.6$, and the Doppler factor is
$\delta=11$.

The difference of $45\arcdeg$ in the position angle of the jet
structure from that measured twelve years previously by Kollgaard et
al.\ (1996) is interesting.  Changes in position angles of component
ejections have been well observed in other sources such as BL Lac, and
have been interpreted as periodic precession of the jet nozzle caused
by binary supermassive black holes by some authors (e.g., Stirling et
al.\ 2003, see also Mutel \& Denn 2005).  
To find out whether or not such variations are occurring
in H\,1426+428 will have to wait for the ejection of more components,
because the two position angles measured twelve years apart are not
enough to place any constraints on such a motion. We note that we do
not detect any significant changes in the position angle of component
C1 itself over the two years of monitoring covered in this paper
(the change in the position angle of C1 of up to $5.4\arcdeg$ in Table~\ref{mfittab}
corresponds to a transverse distance of only about 0.15 mas, about the size of
the 1$\sigma$ positional error bars for this source).

\begin{figure*}[t]
\begin{center}
\includegraphics[scale=0.40]{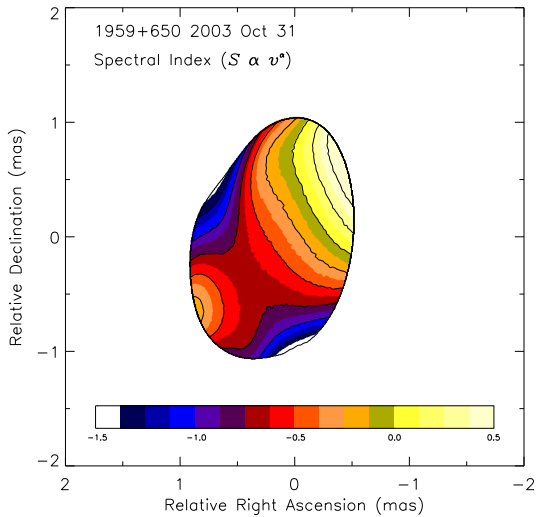}
\end{center}
\caption{Spectral index map of 1ES\,1959+650 between 15 and 22\,GHz, on 2003 Oct 31.
Contours are drawn from spectral indices of $-$1.5 to 0.5 in steps of 0.25.
Spectral index is calculated as $S\propto\nu^{\alpha}$, where $S$ is the flux density,
$\nu$ is the observing frequency, and $\alpha$ is the spectral index. The spectral index has
been plotted at all points where the flux density is at least four times the flux density of the lowest
contour on the images at both frequencies, i.e, at least three contours up on both relevant images.}
\end{figure*}

\subsection{1ES\,1959+650}

Observational background on the $z=0.047$ HBL 1ES\,1959+650 for results
published prior to 2004 has been discussed in Paper~I. Some more
recent high-energy observational results include a spectral analysis
of the 2002 TeV flares, including the ``orphan'' TeV flare in 2002
June (a TeV flare that occurred without a corresponding X-ray flare)
by Daniel et al.\ (2005), and an analysis of low-level TeV emission
detected by the MAGIC telescope in 2004 by Albert et al.\ (2006).
Multi-wavelength campaigns on 1ES\,1959+650 from 2002 and 2003 are
described by Krawczynski et al.\ (2004) and Gutierrez et al.\ (2006),
respectively.  Both of these papers model the multi-wavelength SED of
1ES\,1959+650 with a simple one-zone SSC model with a Doppler beaming
factor of $\delta=20$.  Krawczynski et al.\ (2004) also discuss the
difficulty of modeling the orphan flare mentioned above with
conventional one-zone SSC models, and they suggest several
alternatives, including multiple-component SSC models, external
Compton models, and high-energy proton models.  A specific model for
producing orphan flares from relativistic protons is described by
B\"{o}ttcher (2005), but may require unreasonably high jet powers
(B\"{o}ttcher 2005, erratum).  A multiple-component leptonic model
that succeeds in producing orphan flares is described by Kusunose and
Takahara (2006), using an inhomogeneous conical jet geometry, with the
line-of-sight inside the opening angle of the jet.

On the parsec scale, the 5\,GHz VLBA images of 1ES\,1959+650 by Rector,
Gabuzda, \& Stocke (2003) and Bondi et al.\ (2001, 2004) all show a
diffuse jet with a broad ($\sim55\arcdeg$) opening angle extending 20
mas north of the core along a position angle of $\approx-5\arcdeg$.
VLA images by Rector et al.\ (2003) and Giroletti et al.\ (2004a) show
faint extended flux to both the north (P.A. $\approx-5\arcdeg$) and
south (P.A. $\approx175\arcdeg$) of the core, suggesting negligible
Doppler boosting by the time the jet has reached the arcsecond scale.
Giroletti et al.\ (2006) estimate a viewing angle of
$\theta\sim20\arcdeg$ and a Doppler factor of $\delta\sim3$ on the
parsec scale from the available radio data, similar to their result
for H\,1426+428 discussed above.

Our three epoch series of 15\,GHz VLBA images of 1ES\,1959+650 from
2003-2004 is shown in Figure~3, and the series of 22\,GHz images from
the same epochs is shown in Figure~4.  Both sets of images are shown
restored with both uniform and natural weighting.  Parameters of these
images are given in Table~\ref{imtab}. At the redshift of 1ES\,1959+650
($z=0.047$), the linear scale of the images is 0.9 pc mas$^{-1}$.
These images can then be compared with our previous sequence of three
15\,GHz VLBA images from 2000, as shown in Paper~I.  The 15 and
22\,GHz images show a similar parsec-scale morphology, that is also
similar to that seen for this source in Paper~I, but which is markedly
different from that seen in the lower resolution 5\,GHz VLBA images by
Rector et al.\ (2003) and Bondi et al.\ (2001, 2004).  Figures~3 and 4
show a brighter, compact northwestern feature, with a short extension
to the southeast along a position angle of $\approx125\arcdeg$ ---
this is particularly evident in the uniformly weighted 22\,GHz images.
This extension matches well the location of component C2 seen in Paper~I.  
The diffuse northern jet seen in the 5\,GHz VLBA images is only
faintly present in the naturally-weighted 15\,GHz images. If we apply
a taper to our 15\,GHz VLBA data, then we better detect the diffuse
northern emission.  Figure~5 shows a tapered image from our 15\,GHz
observation on 2004 Feb 8; parts of the northern jet are clearly
visible, although at relatively low significance.

The morphology is intriguing because of the extreme misalignment
between the parsec-scale structures seen at 5\,GHz on the one hand,
and 15 and 22\,GHz on the other.  If the northwestern component on our
images is the core (assumed in Paper~I), then the jet starts to the
southeast with a position angle of $\approx125\arcdeg$ at 0.5 mas from
the core, before bending to a position angle of $\approx-5\arcdeg$ at
about 5 mas from the core. Such an extreme parsec-scale misalignment
would most likely be due to a smaller intrinsic bend, amplified by
projection effects caused by a small angle to the line-of-sight.
Alternatively, the fainter southeastern feature could be the core,
with the northwestern feature representing a brighter jet component,
as has been seen in 4C~39.25 (e.g., Alberdi et al.\ 1993).
Definitively identifying the core requires spectral information, which
is obtained here from the dual-frequency 15 and 22\,GHz observations.
The core is expected to have an inverted or nearly flat spectral index
due to synchrotron self-absorption, while the jet components are
expected to have optically thin spectral indices.

A parsec-scale spectral index map of 1ES\,1959+650 for the 2003 Oct 31
epoch is shown in Figure~6.  The other two epochs yield similar
spectral index maps. The map was constructed by restoring the 22 and
15\,GHz images from that date with the same beam (using the average of
the naturally-weighted beams at each frequency), and calculating the
spectral index at each pixel.  The sign convention used is
$S\propto\nu^{\alpha}$, where $S$ is the flux density, $\nu$ is the
observing frequency, and $\alpha$ is the spectral index.  The
northwest corner of the source has the flattest spectral index, with
the spectral index steepening to the southeast, with a possible small
rise again at the location of the stationary component C2. The
spectral index map thus identifies the core as the brighter component
at the northwest end of the source, as was assumed in Paper~I, a
result confirmed by the dual-frequency model fitting described below.
This leaves a jet with an $\approx130\arcdeg$ apparent bend from the
southeast to the north between 15\,GHz and 5\,GHz VLBI scales as the
most viable interpretation of the parsec-scale morphology.

The source visibilities are well modeled by two circular Gaussians at
both frequencies; the parameters of these models are given in
Table~\ref{mfittab}.  The measured brightness temperature of the core
component is a few times $10^{10}$~K, and it is partially resolved
only at the final epoch at both frequencies, with a brightness
temperature upper limit of $6\times10^{10}$~K at both frequencies at
that epoch.  This is the same as the brightness temperature upper
limit obtained for this source in Paper~I.  The jet component is
blended with the core component on the CLEAN images, but because of
its relatively high flux density relative to the core, it is highly
significant in the model fitting.  The location of this component
matches well the expected location of the component labeled C2 in
Paper~I, so this component is subsequently referred to as C2 (but note
that the same caveats discussed in the section on H\,1426+428 about the
interpretation of model fitting data over widely spaced epochs still
apply).  The component labeled C1 in Paper~I that appeared south of C2
at a separation of about 0.8 mas from the core and a position angle of
160$\arcdeg$ is no longer detected, presumably it has faded between
2001 and 2003. Interestingly, a larger and diffuse Gaussian component
almost on top of the core is suggested by the model fitting, but not
at a very significant level (so it is not included in Table~\ref{mfittab}),
but we speculate that this may represent the older component C1
crossing almost over the core as the jet bends to the north.

\begin{figure*}[p]
\begin{center}
\includegraphics[scale=0.40,angle=90.0]{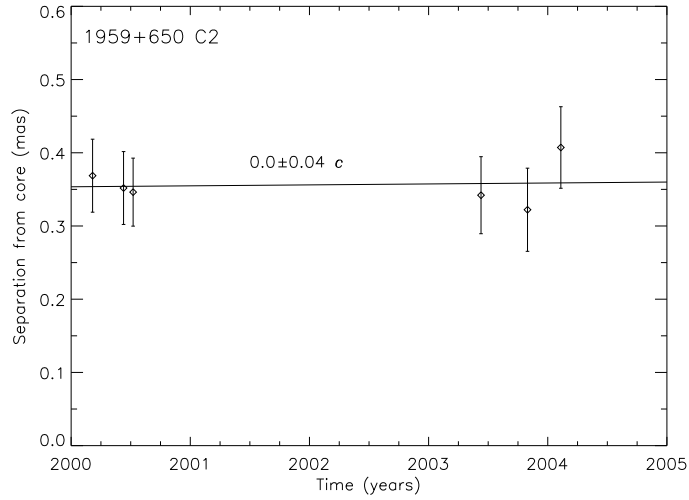}
\caption{Distance from the core of the center of Gaussian component C2 in 1ES\,1959+650 as
a function of time. The line is the least-squares fit to outward motion with constant speed.}
\end{center}
\end{figure*}

\begin{figure*}
\begin{center}
\includegraphics[scale=0.425]{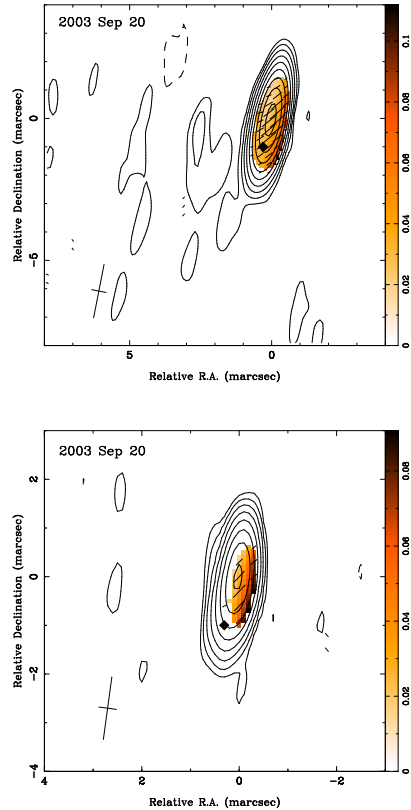}
\end{center}
\caption{VLBA images of PKS\,2155$-$304 at 15.4\,GHz.
The top image is shown using natural weighting,
the bottom is shown using uniform weighting.
Numerical parameters of the images are given in Table~\ref{imtab}.
The location of the center of the circular Gaussian (excluding the Gaussian representing the core)
that was fit to the visibilities is marked with a diamond.
The tick marks show the magnitude of the polarized flux (with a scale of 0.2 mas mJy$^{-1}$
in the top image, and 0.1 mas mJy$^{-1}$ in the bottom image) and the direction of the EVPA.
Tick marks are drawn at pixels where the polarized flux is greater 
than three times the rms noise in the polarization image.
The colors show the fractional polarization, with the scale indicated to the right of the images.}
\end{figure*}

\begin{figure*}[t]
\begin{center}
\includegraphics[scale=0.40,angle=0.0]{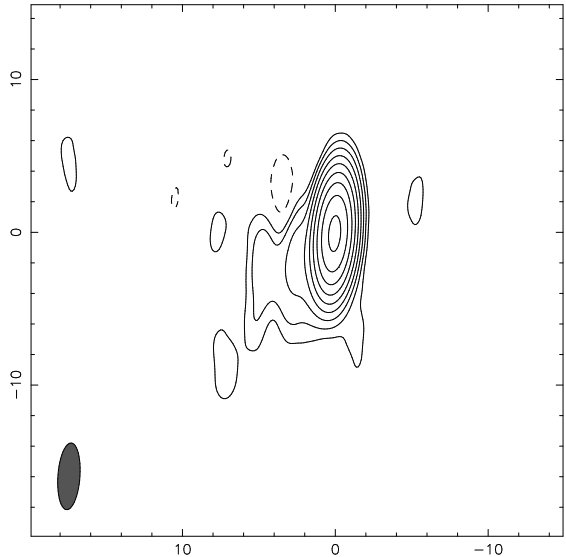}
\end{center}
\caption{Tapered 15\,GHz VLBA image of PKS\,2155$-$304 from 2003 Sep 20.
The restoring beam is 4.38 by 1.44 mas at a position angle of $-4\arcdeg$.
The lowest contour is drawn at a flux density of 0.45\,mJy beam$^{-1}$
(three times the rms noise level). Successive contours are each a factor of 2 higher.
The peak flux density is 140\,mJy beam$^{-1}$.}
\end{figure*}

The spectral indices of the two components can also be calculated from
these dual-frequency model fits.  We average the spectral indices
calculated at all three epochs, and obtain an average spectral index
for the core of $0.04\pm0.18$, and an average spectral index for C2 of
$-0.85\pm0.18$, confirming the results obtained from the spectral
index map. The errors on the spectral indices were calculated from an
estimated 7\% flux error at 15\,GHz and an estimated 9\% flux error at
22\,GHz, where these estimated flux errors were obtained from the
average correction applied by the {\em gscale} amplitude
self-calibration routine in Difmap to the antenna amplitude scales.
The average separation of component C2 from the core is slightly
larger at 22\,GHz than its average separation at 15\,GHz (by
$0.05\pm0.04$ mas), a result expected from inhomogeneous models of the
radio core, where the `core' lies farther back along the jet at higher
frequencies (e.g., K\"{o}nigl 1981).  This measured
frequency-dependent separation was compensated for in the spectral
index map shown in Figure~6 by applying a relative shift in the proper
direction to align component C2 between the two frequencies.

A linear fit to the separation of the jet component C2 from the core
vs. time from the six total 15.4\,GHz images (three from this paper,
three from Paper~I) is shown in Figure~7. The fit yields a measured
apparent speed for this component of $0.00c\pm0.04c$, compared to a
measured apparent speed of $-0.21c\pm0.61c$ for this same component
based on only the three epochs from 2000 that were analyzed in Paper~I.  
The addition of the three new epochs thus lowers the upper limit
on the apparent speed of this component by a factor of ten, from
$0.40c$ to $0.04c$.  Because of the low upper limit on the apparent
speed, this component most likely represents a stationary component in
the jet, whose apparent speed is not directly related to the bulk
apparent speed of the jet (see the discussion on stationary and moving
components in $\S$~\ref{disc}).  The flux density of the component is
variable over the six epochs, as might be expected from a stationary
feature with variable jet plasma passing through it.  If only the
three epochs from this paper are used in the fitting, then an apparent
speed of $0.27c\pm0.35c$ is obtained for the 15\,GHz data, and an
apparent speed of $0.59c\pm0.29c$ is obtained from the 22\,GHz data.
We expect that these differences simply represent the uncertainty
inherent in fitting an apparent speed from three epochs spaced over
only half a year, and do not represent an actual acceleration of the
component.

\begin{figure*}[t]
\begin{center}
\includegraphics[scale=0.40,angle=90.0]{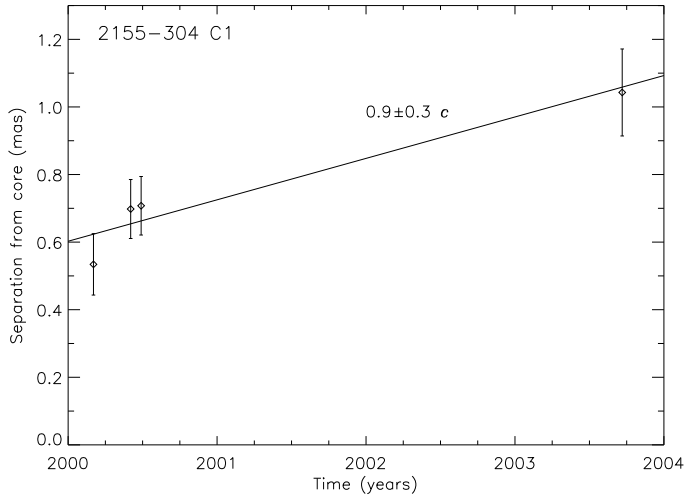}
\caption{Distance from the core of the center of Gaussian component C1 in PKS\,2155$-$304 as
a function of time. The line is the least-squares fit to outward motion with constant speed.}
\end{center}
\end{figure*}

\subsection{PKS\,2155$-$304}

The $z=0.116$ HBL PKS\,2155$-$304 is a prototypical object of this
class, and has been extensively observed at high energies. In
particular, since 2002, its very high-energy gamma-ray flux has been
monitored with the H.E.S.S. array of atmospheric-Cherenkov telescopes
(Aharonian et al.\ 2005a).  A multi-wavelength campaign undertaken in
2003 is described by Aharonian et al.\ (2005b).  These authors model
the SED of this object during a low state, and fit the spectrum with
several SSC model variations with Doppler boosting factors ranging
between 25 and 50.  Of particular note is the recent dramatic flaring
activity at TeV energies on 28 July 2006 (Aharonian et al.\ 2007;
Sakamoto et al.\ 2007).
Bursts varying on timescales of 200 s were observed during this
outburst, requiring Doppler boosting factors greater than 100 if the
emission region has a size comparable to the Schwarzschild radius of a
$\sim10^{9}M_{\odot}$ black hole (Aharonian et al.\ 2007).

Probably because of its declination, this source has not been
well-observed in the radio.  A series of images showing the
large-scale radio structure of PKS\,2155$-$304 is shown by
Laurent-Muehleisen et al.\ (1993).  The highest-resolution VLA image
presented by these authors shows a knot nearly 180$\arcdeg$ misaligned
from the VLBA jet seen in Paper~I; the lower-resolution VLA images by
these same authors show an extended halo of emission around the core.
This source has not been well-observed with VLBI, and the only other
published VLBI images known to us are the three-epoch set from Paper~I.

Our new 15\,GHz VLBA images of PKS\,2155$-$304 from 2003 Sep 20 are
shown in Figure~8.  The linear resolution of these images at the
distance of PKS\,2155$-$304 
is approximately 2.1\,pc\,mas$^{-1}$.  Parameters of these images are given in
Table~\ref{imtab}.  The tick marks on the images indicate the
magnitude of the polarized flux and the direction of the EVPA, and the
color scale indicates the fractional polarization.  These are the
first parsec-scale polarization images of this object.  The
total-intensity morphology is consistent with that seen in Paper~I,
with a compact jet component to the southeast of the core (here at a
position angle of $\approx160\arcdeg$), and a diffuse jet with a broad
opening angle extending to the east.  This diffuse eastern jet is more
easily seen in the tapered image from the same dataset shown in
Figure~9.

Polarized flux is significantly detected at the location of the core
component in both images in Figure~8. The percentage polarization at
the model-fit position of the core (the origin) is 2.9\%, and the
electric vector position angle is $131\arcdeg$, about $30\arcdeg$
misaligned from the innermost jet position angle.  The fractional
polarization increases to the west of the model-fit core position, to
peak values of 8--10\% (see the color scale images in Figure~8).
Because the observations are at only a single frequency, no
corrections have been made for Faraday rotation.  However, the
measured rotation measures in BL Lac objects are typically a few
hundred rad m$^{-2}$ or less (Zavala \& Taylor 2003, 2004), resulting
in an expected error of only about 10$\arcdeg$ or so to the EVPA at
15\,GHz.  The foreground integrated rotation measure in this region is
low, resulting in a negligible correction of under 1$\arcdeg$
(Simard-Normandin, Kronberg, \& Button 1981).

As in Paper~I, the visibilities are well fit by two circular
Gaussians, consisting of the compact core and a single jet component
to the southeast of the core.  Parameters of the model fit are given
in Table~\ref{mfittab}.  The core component is partially resolved with
a best-fit size of 0.23 mas and a lower limit to its size of 0.19 mas,
corresponding to an upper limit to its brightness temperature of
$2\times10^{10}$~K.  The single jet component is slightly farther from
the core than the jet component seen in Paper~I, and we identify it as
the same component, named C1 in Paper~I.  As mentioned above,
identifying components across a several-year gap can be risky, but the
structure of this source is simple, consisting of only a single jet
component, so we consider this identification to be 
likely.
A fit
to the outward motion of this component using the combined four-epoch
dataset from this paper and Paper~I is shown in Figure~10.  The fitted
apparent speed is $0.93c\pm0.31c$. The measurement error on the
apparent speed has been reduced by a factor of nearly ten over the
apparent speed given in Paper~I of $4.37c\pm2.88c$. The new apparent
speed measurement is considerably lower than the previous value, but
given the large associated error on the previous result, they are
statistically consistent at the 1.2$\sigma$ level.  The component
faded over the three years since it was first detected, from about 50
mJy in early 2000 (see Paper~I) to about 15\,mJy in 2003. It has also
turned slightly to the south, moving from a position angle of
$\approx150\arcdeg$ in 2000 to $\approx160\arcdeg$ in 2003.

\begin{table*}[t]
\begin{center}
\caption{Apparent Component Speeds in TeV HBLs}
\label{speedtab}
\begin{tabular}{l c c c c} \tableline \tableline 
& & & & Fastest \\
& & Apparent Speed$^{a}$ & & Apparent Speed \\ 
Source & Comp. & (multiples of $c$)  & Ref. & (multiples of $c$) \\ \tableline 
Mkn~421        & C4    & $0.09\pm0.07$  & 1 & $0.10\pm0.02$ \\ 
               & C4a   & $-0.06\pm0.09$ & 1 &               \\ 
               & C5    & $0.10\pm0.02$  & 1 &               \\ 
               & C6    & $0.03\pm0.03$  & 1 &               \\ 
               & C7    & $0.06\pm0.01$  & 1 &               \\ 
H\,1426+428     & C1    & $2.09\pm0.53$  & 4 & $2.09\pm0.53$ \\
Mkn~501        & C1    & $0.05\pm0.18$  & 2 & $0.54\pm0.14$ \\ 
               & C2    & $0.54\pm0.14$  & 2 &               \\ 
               & C3    & $0.26\pm0.11$  & 2 &               \\ 
               & C4    & $-0.02\pm0.06$ & 2 &               \\ 
1ES\,1959+650   & C1    & $-0.11\pm0.79$ & 3 & $0.00\pm0.04$ \\ 
               & C2    & $0.00\pm0.04$  & 4 &               \\ 
PKS\,2155$-$304 & C1    & $0.93\pm0.31$  & 4 & $0.93\pm0.31$ \\ 
1ES\,2344+514   & C1    & $1.15\pm0.46$  & 3 & $1.15\pm0.46$ \\ 
               & C2    & $0.46\pm0.43$  & 3 &               \\ 
               & C3    & $-0.19\pm0.40$ & 3 &               \\ \tableline
\end{tabular}
\end{center}
$a$: for $H_{0}=71$ km s$^{-1}$ Mpc$^{-1}$, $\Omega_{m}=0.27$, and $\Omega_{\Lambda}=0.73$.\\
References. --- (1) Piner \& Edwards (2005). (2) Edwards \& Piner (2002)
with modified cosmological parameters. (3) Piner \& Edwards (2004) (Paper I). (4) this paper.
\end{table*}

\section{Discussion}
\label{disc}

A major data product resulting from the high-resolution multi-epoch
VLBA monitoring are the apparent jet speeds.  In Table~\ref{speedtab}
we give a table of all apparent speeds that we have measured in TeV
blazars over the course of our monitoring program. This table is meant
to update Table~3 of Paper~I, and it contains our results for Mkn~421
from Piner \& Edwards (2005), as well as the new apparent speeds for
H\,1426+428, 1ES\,1959+650, and PKS 2155$-$304 measured in this paper.
As noted in Paper~I, the apparent pattern speeds measured in the TeV
blazars are considerably slower than those measured in sources
selected for their compact radio emission (Kellermann et al.\ 2004), or
for their GeV gamma-ray emission (Jorstad et al.\ 2001).  The updated
apparent speed measurements added in this paper strengthen the
statistical significance of these conclusions.  Applying the
Kolmogorov-Smirnov test to the apparent speed distribution in
Table~\ref{speedtab} compared to the apparent speed distributions in
Kellermann et al.\ (2004) and Jorstad et al.\ (2001) shows that the TeV
blazars have slower pattern speeds with $>99.75\%$ confidence compared
to the compact radio sources of Kellermann et al.\ (2004), and with
$>99.99\%$ confidence compared to the EGRET blazars of Jorstad et al.
(2001).

VLBA observations of the large 2 cm survey and MOJAVE survey have
shown that powerful jets form VLBI patterns that move at apparent
speeds ranging from zero up to the apparent bulk speed of the jet
(Lister 2006; Cohen et al.\ 2007). These authors conclude that the peak
apparent pattern speed measured in a jet from VLBI monitoring is then
a good indicator of the apparent bulk speed of the flow.  In the final
column of Table~\ref{speedtab}, we therefore list the fastest pattern
speed observed for each of these TeV blazars during our monitoring.
All of these fastest speeds are below about 2$c$, and four of the six
are subluminal. One of the following two possibilities must then apply
to the apparent pattern speeds in the parsec-scale jets of the TeV
HBLs:
\begin{enumerate}
\item{The pattern speeds are related to the bulk apparent speeds as described above, and the
bulk apparent speeds are slower on the parsec scale in the TeV HBLs compared to the more powerful sources.}
\item{A much larger fraction (approaching 100\%)
of the HBLs have pattern speeds that are unrelated to their bulk apparent speeds
than do the more powerful sources.}
\end{enumerate}
Either of these two possibilities implies a fundamental difference between the parsec-scale jets of HBLs and
the more powerful sources, and we discuss each possibility in turn:

\subsection{Pattern speeds related to bulk speeds}

If the fastest pattern speeds are an indication of the bulk apparent
speed, as they are for the MOJAVE sources, then we have a typical bulk
apparent speed of about 1$c$ in the parsec-scale jets of the TeV
blazars (the mean and median of the final column in
Table~\ref{speedtab} are both about 1$c$).  The bulk apparent speed is
a function of the bulk Lorentz factor and the viewing angle, as is the
Doppler factor (see equations in footnote 3).  If the high Doppler
factors implied by the high-energy observations are combined with the
slow apparent speeds then very small viewing angles are obtained (many
under one degree), that imply unreasonable numbers of parent objects
(Henri \& Saug\'{e} 2006).  A statistically reasonable solution is
then to adopt a lower Doppler factor (and therefore Lorentz factor) in
the parsec-scale radio-emitting region than in the gamma-ray emitting
region.  This argument was discussed in more detail in Paper~I.

A lower Doppler factor and bulk Lorentz factor for the parsec-scale
radio emission in HBLs is actually suggested by many observations
other than the apparent pattern speeds. Each of the observables
described below depends on the bulk Doppler factor of the jet, not on
a pattern speed:
\begin{description}
\item[a.]{The TeV blazars are significantly less variable in the radio
    band than are the EGRET blazars (Aller, Aller, \& Hughes 2006).}
\item[b.]{The TeV blazars have low radio core brightness temperatures
    (see Table~\ref{mfittab}) that do not require invoking
    relativistic Doppler factors to reduce them below possible
    intrinsic limits.  Typical high-power sources have much higher
    brightness temperatures (Kovalev et al.\ 2005; Dodson et al.\
    2007).}
\item[c.]{The measured radio powers of the cores suggest Lorentz
    factors for HBLs of around $\Gamma=3$, when this power is compared
    to that expected from the total power at low frequency (Giroletti
    et al.\ 2004a).}
\item[d.]{The calculated inverse Compton Doppler factor from the VLBI
    cores of HBLs gives a mean lower limit to the Doppler factor of
    $\delta\gtrsim4$, and a mean lower limit to the Lorentz factor of
    $\Gamma\gtrsim2$ (see Table~7 in Giroletti et al.\ 2004a). Thus, high
    Lorentz factors are not required to reduce the predicted X-ray
    emission from the VLBI cores below the observed values.}
\item[e.]{Optical and radio luminosities of HBLs could not be unified
    with their FR~I parent population using only transverse jet
    velocity structures by Chiaberge et al.\ (2000), but the required
    properties could be reproduced by assuming that the radio emitting
    region in HBLs is less beamed than the optical one, as could be
    expected if the jet decelerates after the higher energy emitting
    zone (Chiaberge et al.\ 2000).}
\item[f.]{The VLBI jets of the TeV blazars assume a plume-like
    morphology beyond a few mas from the core, in contrast to
    high-power jets that may remain well-collimated to large
    distances.  This transition to a plume-like morphology has been
    interpreted as evidence for entrainment induced deceleration in
    TeV blazars by Bicknell et al.\ (2005).}
\end{description}

These six observational properties all imply relatively low bulk Doppler
factors and Lorentz factors in these parsec-scale jets.  It is clear
that radio observations consistently yield Doppler factors and Lorentz
factors for the TeV blazars (and HBLs in general) about an order of
magnitude lower than the high-energy observations.  On the other hand,
arguing for a higher Lorentz factor is the fact that no counterjets
have been observed in any of the TeV HBLs. Our observations put the
limit on the jet to counterjet ratio $J$ at $J\gtrsim100$ for most of
the sources. The tightest limits on $J$ in a TeV blazar come from
High-Sensitivity Array (HSA) observations of Mkn\,501 by Giovannini,
Giroletti, \& Taylor (2007), which constrain $\beta\cos\theta>0.92$,
corresponding to $\Gamma\gtrsim3$ for viewing angles of a few degrees.
Thus, a parsec-scale Lorentz factor in the radio-emitting region of
$\Gamma\sim3$ seems consistent with the six points given above, and is
just consistent with the observed jet-to-counterjet brightness ratios.
A Lorentz factor of $\Gamma\sim3$ is also consistent with the fastest
apparent speeds given in Table~\ref{speedtab} (except for the very
slow speeds in Mkn~421 and 1ES\,1959+650), and yields reasonable
viewing angles of 2$\arcdeg$ to 8$\arcdeg$ for the apparent speeds
between 0.5$c$ and 2$c$.  It is not surprising that these viewing
angles may be substantially less than 1/$\Gamma$, since these sources
were selected based on their TeV emission which may have a narrower
beaming cone.

The crucial questions then become: where is the parsec-scale radio
emission located relative to the site of the gamma-ray emission, and
what causes its lower Lorentz factor?  One possibility is that the
parsec-scale radio emission (at light-year scales) is located
downstream of the gamma-ray emission (at light-day scales), and that
the jet decelerates as it moves out.  Georganopoulos \& Kazanas (2003)
considered such a jet that decelerates along its length, and showed
that the required values of the Doppler factor in the inner jet are
not as high as in the homogeneous models discussed above, because the
fast inner portion of the jet can more efficiently upscatter
blueshifted synchrotron photons from the slower outer portion.  Thus,
allowing an inhomogeneous model reduces some of the apparent conflict.
Wang et al.\ (2004) modeled a decelerating jet where the dissipated
kinetic energy of the jet is transferred to electron acceleration,
they then used the resulting radiation from the accelerated electrons
to successfully model the SED of several TeV blazars.  Bicknell et al.
(2005) have also modeled parsec-scale TeV blazar jet deceleration that
they claim physically reflects a conversion of kinetic luminosity into
enthalpy flux, and concluded that such deceleration can be physically
consistent with the conservation of energy and momentum in this
region.

A second possibility is that the jet has an inhomogeneous structure
transverse to the jet axis, consisting of a fast `spine' which
dominates the high-energy emission, and a slower `layer', which
dominates at lower frequencies, although such a structure is not
mutually exclusive with the decelerating jet discussed above.  For
example, Ghisellini et al.\ (2005) proposed a jet with an initial {\em
  transverse} velocity structure consisting of a fast spine and a slow
layer, and by taking into account the inverse Compton radiation each
portion produced from the seed radiation coming from the other
portion, they found that the resulting anisotropic inverse Compton
emission from the spine may serve to decelerate it.  Henri \&
Saug\'{e} (2006) propose a two-flow model, consisting of a mildly
relativistic outer MHD jet, and a central pair-plasma jet accelerated
by the so-called `Compton rocket' effect to higher velocities of order
$\Gamma\sim3$. Regardless of their origin, such transverse structures
may show an observational signature in the VLBI images of jets that
are resolved in the transverse direction, in the form of
limb-brightening or limb-darkening, according to which region is
dominating the radio emission. Observational signatures of
limb-brightening have been claimed in the VLBI images of the TeV
sources Mkn~501 (Giroletti et al.\ 2004b), Mkn~421 (Giroletti et al.
2006), and M~87 (Kovalev et al.\ 2007). We also found some
observational evidence of limb-brightening in the Mkn~421 jet (Piner
\& Edwards 2005), but it was present in only some of the transverse
slices, suggesting a more complex transverse emission pattern on the
scales that we investigated.  The jets of the three fainter (in the
radio) TeV sources considered in this paper are not detected with
sufficient resolution or dynamic range in the VLBA images to discern
structure transverse to the jet axis; those observations must be left
to the brighter sources (or to much more sensitive images of these
sources).

A third possibility, that the gamma-ray emission is co-spatial with
the parsec-scale jet knots, was suggested by observations of knot
HST-1 in M~87 (recently reviewed by Harris et al.\ 2007).  That knot,
lying at a distance of $\sim100$ pc from the core of M~87, would lie
at angular separations of a few mas from the core in the $z\sim0.1$
TeV blazars, taking into account likely projection effects in jets
with viewing angles of a few degrees. That would put features like
HST-1 in M~87 co-spatial with the VLBI model components in the jets of
the TeV blazars. If the high-energy emission is occurring in these
components, that makes the different Lorentz factors and Doppler
factors derived in the radio and the gamma-ray particularly
challenging to explain.

\subsection{Pattern speeds unrelated to bulk speeds} 

Many sources in large VLBI surveys display a mix of stationary and
moving components --- but what we mean here is that even the {\em
  fastest} measured apparent pattern speed in a source may still be
much less than the bulk apparent speed.  The study of the 2 cm and
MOJAVE survey apparent speeds (Cohen et al.\ 2007) found that a small
fraction (about 25\%) of the powerful sources showed {\em only} slow
apparent pattern speeds that were unrelated to their bulk apparent
speeds based on statistical arguments (otherwise they would have
unreasonably small angles to the line-of-sight, or unreasonably high
intrinsic luminosities). If the bulk Lorentz factor is indeed high in
the parsec-scale jets of the TeV HBLs, then the fraction of these
sources that have apparent pattern speeds much less than their
apparent bulk speed is closer to 100\%.

Is there then some plausible physical difference in the high and
low-power jets that could cause the high-power jets to form patterns
that move at the bulk speed, while the low-power jets do not?  A
common model for particle acceleration and the formation of VLBI
components is the shocked-jet model (e.g., Spada et al.\ 2001). If such
rapidly moving shocks were forming and propagating to the parsec-scale in the TeV
HBLs, then we might expect them to be visible in the VLBA monitoring,
and they are not.  Krawczynski (2007) has suggested a model where
particle acceleration in the low-power sources occurs through
acceleration of parallel electron-positron or electron-proton beams
rather than through shock acceleration.  If a source does not form
strong moving shocks, then its more continuous jet may be modeled by a
series of stationary Gaussians, even though the plasma is moving
relativistically.  Other features that are due to standing shocks, or
to points where the jet bends toward the line-of-sight, may also
appear in jets as stationary regions of locally-enhanced emission, as
they do in high-power sources.

Another possible physical difference between the high and low-power
jets that may cause such a difference in pattern speeds is the opening
angle of the (presumably conical) jet.  This possibility is explored
in the series of papers by Gopal-Krishna et al.\ (2004, 2006, 2007).
According to these models, if the low-power sources have jets with
considerably larger opening angles (Gopal-Krishna et al.\ 2007), then
what is seen in the VLBI observations is a set of observables that
have been unintentionally averaged over different viewing angles, with
the small viewing-angle portion of the jet being the most Doppler
boosted, and contributing the most to the weighted average. In that
case, many sources may show slow apparent speeds typical of being
viewed almost end-on, while the apparent speed {\em on the axis} of
the jet is actually much higher.  However, in all of the versions of
this model that produce subluminal effective apparent speeds, the
effective Doppler factor remains quite high ($\delta_{eff}\gtrsim20$,
see Figures~3 and 4 of Gopal-Krishna et al.\ 2007), so that this model
cannot easily explain the relatively low Doppler factors measured for
the TeV blazars in the radio as discussed in points {\bf a} through
{\bf f} in the previous subsection (in fact, it would predict high
values for the radio variability, brightness temperature, core
dominance, and inverse Compton Doppler factors). 

Whether or not the
opening angle has a significant effect on the observables hinges on
the actual value of the opening angle --- the effect is small for
1$\arcdeg$ opening angles and quite pronounced at 5$\arcdeg$ opening
angles; but, perversely, measurements of blazar jet opening angles are
of little use in testing the model, because the measurements of the
opening angle are themselves effected by averaging over the true
opening angle (see Figure~3, Gopal-Krishna et al.\ 2006).  However, if
the opening angles are large enough to cause a significant effect, it
would seem that a substantial number of sources may be expected to
have core-halo type morphology on their VLBI images (whenever the
opening angle is larger than the axis viewing angle), so it would be
useful to compare simulated VLBI images from this model with observed
morphologies.  The model also makes specific statistical predictions,
e.g., that a small fraction of TeV blazars should show strongly
superluminal motions (Gopal-Krishna et al. 2006), so that as the
number of TeV blazars increases, it will be possible to test the
observations against probability curves derived from this model.

Based on all of the available observational data, we favor a
combination of possibilities 1 and 2 given at the beginning of this
discussion section.  From all of the other arguments for a low Lorentz
factor for the parsec-scale radio emission given above (points {\bf a}
through {\bf f} in the previous subsection), we consider it likely
that the parsec-scale radio emitting region in the TeV HBLs has a
substantially lower Lorentz factor than is usually derived for the
high-energy gamma-ray emitting region.  A parsec-scale Lorentz factor
of about 3 is indeed consistent with the peak apparent speeds of
0.5$c$ to 2$c$ in Table~\ref{speedtab}, assuming these reflect the
bulk apparent speed on the axis of the jet.  However, the apparent
speeds in two of the TeV blazars (Mkn~421 and 1ES\,1959+650) are so low
that they likely represent a sub-sample of these sources where the
measured pattern speeds are much less than the bulk speed, as was also
found for powerful sources by Cohen et al.\ (2007).  The physical cause
of this Lorentz factor difference between the high and low-frequency
emitting regions (and whether any Lorentz factor gradient occurs along
the jet, transverse to the jet, or both), and the physical cause of
any decoupling of pattern and bulk speeds has not been definitively
determined, but the sample of TeV blazars is growing rapidly, so that
the observational statistics will improve markedly over the next
several years.

\section{Conclusions}
We have presented new multi-epoch VLBI images
for the three TeV blazars H~1426+428, 1ES~1959+650, and PKS~2155$-$304
obtained during the years 2001 to 2004.
The results for H~1426+428 are the first multi-epoch VLBI results to
be presented for this source. The results
for 1ES~1959+650 and PKS~2155$-$304 are combined with earlier 
results for these sources from Paper I to yield a longer time baseline for measurement.
The major observational findings for these three sources are:
\begin{enumerate}
\item{H~1426+428's parsec-scale structure during this time range was well-modeled by a
$\sim 17$\,mJy core and a single $\sim 3$\,mJy jet component at a position angle of
approximately $-25\arcdeg$, with an apparent speed of $2.09c\pm0.53c$.}
\item{1ES~1959+650 consisted of a compact core and a nearby stationary ($0.00c\pm0.04c$)
jet component at a position angle of about $125\arcdeg$ and
a separation of about 0.35 mas at 15 GHz. On larger scales of a few mas the
jet is diffuse and directed to the north, so that this source shows an extreme apparent
misalignment of about $130\arcdeg$ on parsec scales.}
\item{PKS~2155$-$304 was observed with dual-circular polarization at 15 GHz.
The fractional polarization at the
position of the core was 3\%, and the electric vector position angle
was $131\arcdeg$, about $30\arcdeg$ misaligned from the innermost
jet position angle. The measured apparent speed of the single jet component
was $0.93c\pm0.31c$.}
\end{enumerate}

We combined the new apparent speed measurements from this paper
with the apparent speeds measured in TeV blazar jets from our earlier papers to form 
a current set of apparent speed measurements in TeV HBLs (Table~\ref{speedtab}).
The mean peak apparent pattern speed in the jets of the TeV HBLs is about 1$c$.
The statistical result noted in Paper I that the TeV HBLs have significantly slower
apparent jet pattern speeds compared to radio-selected or GeV-selected blazars is strengthened
by the new results of this paper.
The Discussion section ($\S$~\ref{disc}) presented a thorough analysis of these
results in the context of other radio observations and theoretical models for
TeV blazar jets. Conclusions from that analysis were:
\begin{enumerate}
\item{The peak apparent speeds of order 0.5-2$c$ in four of the six 
studied TeV blazars (Table~\ref{speedtab}), when
taken together with other observed radio properties discussed in
$\S$~\ref{disc}, are consistent with
radio jets with bulk Lorentz factors of $\Gamma\sim3$ and viewing angles of 
a few degrees.}
\item{The very slow peak apparent speeds in two of the six studied
TeV blazars in Table~\ref{speedtab} (Mkn~421 and 1ES\,1959+650) are likely pattern
speeds that are unrelated to the bulk apparent speeds of the jets,
which are likely to be similar to those mentioned above.}
\end{enumerate}

\acknowledgments
We acknowledge helpful comments from the referee that improved
the quality of the paper.
The National Radio Astronomy Observatory is a facility of the National
Science Foundation operated under cooperative agreement by Associated
Universities, Inc.  
This research has made use of the NASA/IPAC
Extragalactic Database (NED), which is operated by the Jet Propulsion
Laboratory, California Institute of Technology, under contract with
the National Aeronautics and Space Administration.  This work was
supported by the National Science Foundation under Grant Nos. 0305475
and 0707523.

{\it Facilities:} \facility{VLBA ()}

\end{document}